\documentclass[
showpacs,floatfix,amsmath,amssymb,aps,prstab,reprint,superscriptaddress]{revtex4-1}
\usepackage{graphicx,color}

\newcommand{\ud}{\,\mathrm{d}}
\begin{document}

%\preprint{}

\title{Heating of Micro-protrusions in Accelerating Structures}
\thanks{This work was supported by the Office of High Energy Physics of the U.S. Department of Energy.}


\author{A. C. Keser}
\email[]{akeser@umd.edu}
\author{T.M. Antonsen}
\author{G.S.Nusinovich}
\author{D.G. Kashyn}
\affiliation{Institute for Research in Electronics and Applied Physics, University of Maryland, College Park, Maryland 20742-2511, USA}
\author{K. L. Jensen}
\affiliation{Naval Research Laboratory, Washington D.C. 20375-5347, USA}
%\thanks{}


\date{\today}

\begin{abstract}
The thermal and field emission of electrons from protrusions on metal surfaces is a possible limiting factor on the performance and operation of high-gradient room temperature accelerator structures.
We present here the results of extensive numerical simulations of electrical and thermal behavior of protrusions.
We unify the thermal and field emission in the same numerical framework, describe bounds for the emission current and geometric enhancement,
then we calculate the Nottingham and Joule heating terms and solve the heat equation to characterize the thermal evolution of emitters under RF electric field.
Our findings suggest that, heating is entirely due to the Nottingham effect, that thermal runaway scenarios are not likely, and that
high RF frequency causes smaller swings in temperature and cooler tips. 
We build a phenomenological model to account for the effect of space charge and show that space charge eliminates 
the possibility of tip melting, although near melting temperatures  reached.
\end{abstract}

%\keywords{}

\maketitle
\section{Introduction} \label{sec:Introduction}

The thermal and field emission of electrons from metal surfaces is a limiting factor on the performance and operation of high-gradient room temperature accelerator structures\cite{WangLoew,Grudiev}.
Electron emission results in dark current and has a number of undesirable effects including: parasitic absorption of RF energy, and excessive heating and radiation\cite{WangLoew}. 
Moreover, it is believed to be the precursor to breakdown along with local outgassing and plasma formation\cite{WangLoew,Bonin}. 

Field emission is traditionally described by the Fowler-Nordheim equation\cite{FN}. 
Through the use of this formula, the enhancement of the electric field on a surface can be predicted from experimental measurements of dark current. 
Typical values of 'effective' field enhancement factor measured by this method are in the range of 40 to 100\cite{WangLoew}.
However, the modeling of underlying physical processes that contribute to dark current is a long standing unsolved problem. 

In practice, metal surfaces are not perfectly flat and clean. 
The microscopic imperfections cause a large variation in the local surface electric field. 
Since the field emission is an exponentially increasing function of surface field, strongly emitting sites form where the field is enhanced due to local imperfections.
If the imperfections are due to contaminates, oxides and/or adsorbates , then there is variation in the work function as well.
For purposes of this paper, work function variation due to adsorbate or crystal face variation are not considered.
Metallic protrusions are a type of surface imperfection that can cause enhanced field emission and heating. 
Here, we particularly focus on heating and the possibility of thermal runaway and melting of protrusion tips, as these are put forward as precursors to breakdown in the community\cite{Kevin}.
From these considerations we want to model the processes in a micro-protrusion to answer the following questions: 
i) How does the geometry of the protrusion affect the heating and emission?
ii) How does the temperature of the protrusion evolve and is there a possibility of thermal run away, explosion or tip melting?
iii) How does the RF frequency affect the temperature rise?

Field emitting structures are widely analyzed both in the context of dark current, device breakdown and field emitter technology.
Previous studies that inspired our work can be listed as follows. 
Wang and Loew gave a tutorial account of field emission and RF breakdown and showed elementary calculations related to field enhancement and subsequent emission in RF fields\cite{WangLoew}.
Chatterton, in his theoretical study published in 1965, established limits for the applied electric field, by considering various geometries for the protrusions\cite{Chatterton}.
Ancona numerically analyzed failure in molybdenum field emitters taking Nottingham heating into account which lays down the methodology that we followed in this study\cite{ancona},
and finally, Jensen \textit{et. al.} built an analytically tractable model to describe the geometry and heating of micro protrusions and their contribution to dark current\cite{Kevin}.

Our approach to determine the temperature rise in micro-protrusions is to break the problem into several distinct, but interrelated sub-problems. 
These include: i) the determination of the electric field outside the protrusion,
ii) the determination of the rate of electron emission from the surface of the protrusion including an estimate of the effect of the electron space charge on the electric field near the surface,
iii) the determination of the conduction current in the protrusion and 
iv) the determination of the temperature rise due to Joule heating implied by the conduction current and the Nottingham effect implied by the electron emission.

The physical processes governing these problems occur on different time scales and this will motivate a number of simplifying approximations that will allow these problems to be solved sequentially. 
First, because the protrusion is small in comparison to the vacuum wavelength of the RF field in the structure,
fields outside the protrusion will be essentially electrostatic and will oscillate following the RF field. 
We will assume that the electron space charge field is a small perturbation to the applied field (except near the surface of the protrusion) so that the electric field outside the protrusion is essentially a vacuum field. 
Finally, we will assume that the electrical conductivity of the material composing the protrusion is high such that the potential on the surface of the protrusion can be taken to be a constant (zero) in the calculation of the field outside the protrusion.

The rate of electron emission from the surface of the protrusion will be calculated assuming that the surface is locally flat. 
Here it is assumed that the variation of the electric field along the surface occurs on a scale length that is greater than the atomic dimensions that define the emitting regions. 
Further, we assume that the electrons transit through the emitting region rapidly compared with the time scale for variation of the electric field.  
The result will be an emission law that gives the current density of emitted electrons as a function of the local normal component of the vacuum electric field, 
calculated previously, and the local temperature of the surface.
This emission law will be modified to account for shielding of the emitting surface by the electron space charge layer. 
Associated with the emitted current density, there is a heat flux leaving the surface.
The heat flux may be either positive or negative due to the so-called Nottingham effect \cite{Nottingham}.

The electric field and conduction current inside the protrusion will be calculated under the quasi-static assumption that the relaxation time of the conduction current is much shorter than the period of the RF fields.
Essentially, we assume the dimensions of the protrusion are much less than the skin depth.
This will lead to a Laplace equation for the potential where the boundary conditions are that the potential goes to zero deep inside the conductor supporting the protrusion, and the current density emitted from the surface of the protrusion is given by the previously determined emission law.
Finally, the temperature distribution throughout the protrusion will be determined by the heat diffusion equation with the Joule heating of the previously determined conduction current density serving as a source. Here the boundary conditions are that the temperature rise deep inside the conductor supporting the protrusion is small and that there is a prescribed heat flux on the surface of the protrusion that is calculated in parallel with the current emission law.

The remainder of this manuscript is organized as follows. Section~\ref{sec:outside} describes our calculation of the electric field outside the protrusion. Section~\ref{sec:emission} describes the development of the emission law to be applied at the surface of the protrusion. The discussion of our phenomenological accounting for space charge is contained in Appendix. In section~\ref{sec:u_solution} we describe the determination of the current density in the protrusion, and section~\ref{sec:Temp_rise} is devoted to the temperature rise in the protrusion.

\section{Electric field outside the protrusion } \label{sec:outside}

We will assume the protrusion can be considered to be a conically shaped, cylindrically symmetric, piece of conducting material located on top of a flat conducting surface. We will pick the z-axis as the  vertical axis of symmetry. 

The electric field outside the protrusion is determined by Laplace's equation with the following boundary conditions. The potential on the surface of protrusion and on the surface of the supporting conductor is taken to be zero. Far above the protrusion, the potential matches that of a uniform electric field, where $\mathbf{F_o}=F_o\mathbf{\hat{z}}$ represents the time dependent, normal component of the RF electric field that would be present if the surface was flat (i.e. no protrusion).

We will solve this problem using a point charge approach based on the method of images \cite{Kevin}. 
In this approach the potential is represented as the sum of the potential generating the applied field ($\phi = -z F_o$) 
and the potential due to a set of point charges along the z-axis located in the protrusion, 
and their images located along the z-axis in the conductor below the protrusion. 
The charge and its image constitute a dipole.
The number of dipoles in the model is given by the parameter $N$.
We place the first charge at $z_1=a_o$. 
The rest of the charges are placed at steps with sizes decreasing geometrically with a constant factor $r$, 
so that the position of $n^{th}$ charge becomes: $z_n =\sum_{j=1}^n r^{j-1} a_o$ where $1<n<N$.
The image charges are placed at $-z_n$.
As a result we obtain the following formula for the potential:
\begin{equation}
\label{PCM_potential}
\begin{split}
	V_N(\rho,z) = 
F_o a_o
		\bigg\{
			-\frac{z}{a_o} + 
			a_o \sum_{j=1}^N 
			\frac{\lambda_j}
				{
					\sqrt{\rho^2 +(z-z_j)^2}
				}
				- \\
			\frac{\lambda_j}
				{
					\sqrt{\rho^2 +(z+z_j)^2}
				}
		\bigg\}
\end{split}
\end{equation}
To determine the relative strength of charges, $\lambda_n$, we demand that the potential satisfies the following recursion relation: 
for each $n$, $n=1$ to $N$, $\lambda_n$ is determined by $V_n(0,z_{n+1})=0$.
For example, if we have the model with a single dipole , $N=1$, to find $\lambda_1$, we demand $V_1(0,z_2) = F_o a_o \{-z_2/a_o + a_o \lbrack \lambda_1/(z_2-z_1) -  \lambda_1 /(z_2+z_1) \rbrack \} = 0$.
In the case $N=2$, to find $\lambda_1$ and $\lambda_2$, we solve $V_1(0,z_2)=0$ and $V_2(0,z_3)=0$ sequentially.

\begin{figure}[ht]
	\includegraphics[width=0.4\textwidth]{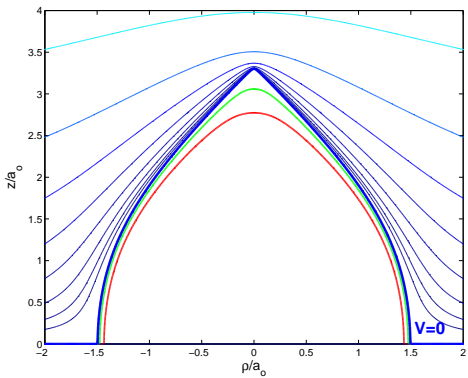}
	\caption{Protrusions constructed with three different values of N. 
			Increasing N results in sharper tips and thus higher enhancement factors.
			The three thick lines with different colors describe the shapes of the following protrusions.
			Red:~$r=0.7,N=4, \beta=8.0$, green:~$r=0.7,N=6,\beta=12.4$, 
			blue:~$r=0.7,N=12,\beta=44.3$
			Also shown is the contour plot of the electric potential obtained by using the point charge model for $r=0.7, N=12$.
			$V=0$ marks the zero equipotential line which constitutes the shape of the protrusion.
	\label{fig:PCMv}}
\end{figure}

\begin{figure}[ht]	
		\includegraphics[width=0.4\textwidth]{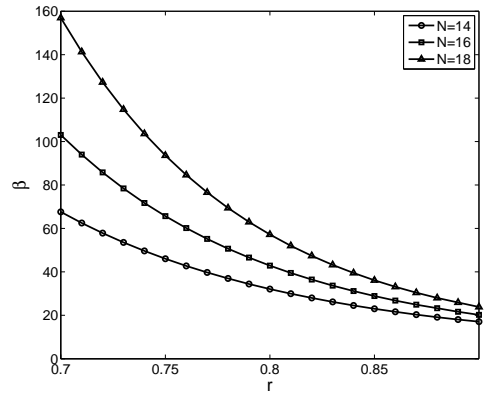}
		\caption{$\beta$; the electric field enhancement factor as a function of ratio $r$ for three different values of point charge number $N$.
	\label{fig:Beta_r}}
\end{figure}

An example is shown in Fig.~\ref{fig:PCMv} for a particular choice of these parameters as described in the caption. 
The thick line shows the surface of the protrusion (zero potential) and the thin lines show level curves for other values of the potential. 
Only the potential outside the protrusion is of interest to us.
We note that the potential curves outside the protrusion, near its tip, are crowded together indicating a strong enhancement of the electric field near the tip.  
Most of the electron emission will occur there.
Fig.~\ref{fig:PCMv} also shows the boundary of protrusions for three different choices of point charge parameters. 
For each protrusion, the field enhancement factor $\beta$, that is defined as the ratio of geometrically enhanced electric field 
at the tip to $F_o$, is calculated. 
It can be seen that varying these parameters, potentials can be constructed for protrusions of differing degrees of pointedness. 
Finally, we plot in Fig.~\ref{fig:Beta_r} the field enhancement factor $\beta$ for different combinations of the point charge model parameters.
\section{Emission Law} \label{sec:emission}
The second step in our procedure is to determine the emitted current density on the surface of the protrusion.
The emission current is a combination of thermal and field emission, which we take to be described by a general thermal field (GTF) emission formula \cite{Kevin}. 

In this treatment the free electrons in the metal face a square potential, which may be modified by external fields and the Schottky effect. 
One then computes the wave function in the presence of this potential and calculates the transmission probability. 
Given the Fermi-Dirac distribution of electrons, the current can be expressed as an integral over energy of the product of supply function and the transmission probability.
\begin{equation}
	J(F,T)=\frac{q}{2\pi \hbar}\int_0^{\infty} D(E)f(E)\ud E
	\label{QMcurrent}
\end{equation}
where $D(E)$ denotes the tunneling or transmission probability as a function of energy and $f(E)$ denotes the electron supply function as a function of energy.
Expressions for $D(E)$ and $f(E)$ can be found in Ref.~\cite{MG}.
The resulting integral in~\eqref{QMcurrent} is a function of the local electric field, $F$, and the temperature, $T$, on the surface.

\begin{figure}[ht]
		\includegraphics[width=0.4\textwidth]{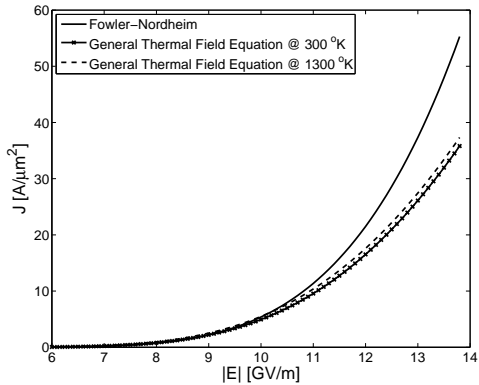}
		\caption{Comparison of Fowler-Nordheim (FN) and General Thermal Field Current Densities for copper work function taken as $\Phi=4.5\:eV$.
	\label{fig:GTF}}
\end{figure}

Equation~\eqref{QMcurrent} describes electron field emission and reduces to the Fowler-Nordheim formula in the low temperature limit\cite{FN}.
In the high temperature low field limit it describes thermionic emission and reduces to the Richardson-Laue-Dushman formula\cite{RLD}.
However, there is a transition region in which both formulas fail to apply.
By using series expansion techniques Murphy and Good proposed a formula valid for the transition region \cite{MG}.
(See Coulombe and Meunier for a comparison of Murphy-Good formula with its predecessors\cite{Coulombe}.)
Finally in 2007 Jensen proposed a general thermal-field emission (GTF) formula for a wide range of temperatures and fields\cite{JensenGTF}.
Here, instead of employing the analytical expressions for the GTF current, we resort to numerical evaluation of Eq.~\eqref{QMcurrent}.
In Fig.~\ref{fig:GTF} we plot the GTF current density obtained from Eq.~\eqref{QMcurrent} at different temperatures for copper, versus peak electric field at the tip.
It should be noted that there is a bound to the maximum electric field that can be treated in this model. 
First, the maximum of the Schottky modified potential barrier should not drop below the Fermi level. 
For example, copper has a work function of $4.5\:eV$, and the barrier maximum is $-\sqrt{e^3 F / (4 \pi \epsilon_o)}$ measured from the vacuum energy\cite{MG}. 
The formula~\eqref{QMcurrent} is a good approximation provided the barrier maximum is above the Fermi level.
This gives a maximum field of around $14\:GV/m$. 
For a background field of $300\:MV/m$ this corresponds to a field enhancement factor of $47$. 
As an example, for $(a_o=1\:\mu m, r=0.78, N=20)$ we obtain a protrusion with $\beta \approx 47$. 
For this protrusion the base radius is $1.59\:\mu m$ and height is $4.58\:\mu m$.
Finally, the value of the electric field at the surface of the protrusion will be modified (reduced)
by the space charge of the emitted electrons. 
When extreme electric fields and very high current densities such as those in Fig.~\ref{fig:GTF} are considered, the space charge effect becomes relevant \cite{Barbour, JensenSC}. 
The creation of a high density layer of electrons around the emitter tip, reduces the vacuum electric field. 
The effect of space charge should be solved consistently, because the space charge reduced electric field
is a function of emission current, hence ultimately the electric field itself.

\begin{figure}[ht]
		\includegraphics[width=0.4\textwidth]{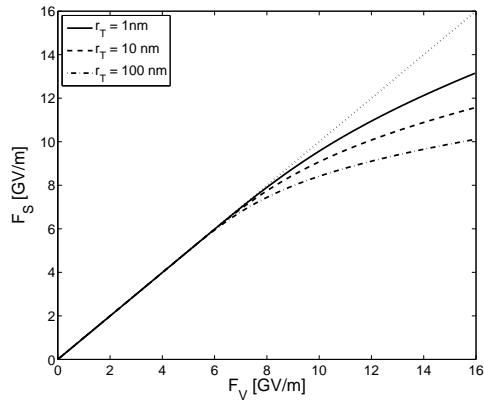}
		\caption{Electric Field Magnitude in the presence of space charge effect $F_S$ compared to its vacuum value $F_V$ for various values of tip radius $r_T$.
		The dotted line shows the case with no space charge.
	\label{fig:SPalpha}}
\end{figure}

In Appendix we describe a model that we use to account for this effect. 
Specifically, we derive a relation between the electric field at the surface in the absence of 
space charge, $F_V$, and the electric field at the surface when space charge is included, $F_S$.
The result of this model is displayed in Fig.~\ref{fig:SPalpha} for protrusions at room temperature
with three different radii of curvature of the tip.
It can be seen that the effect of space charge is to limit the field at the surface, and consequently limit
the emitted current density.
To summarize, as previous studies\cite{MG,Coulombe,Paulini} already established, 
the thermionic and field emissions formulas miscalculate the correct emission 
from a heated field enhancing structure. 
Therefore, we use the general thermal-field (GTF) emission in this study. 

The emission current leads to two mechanisms of heating, namely the Joule heating in the bulk of the protrusion and the Nottingham effect that acts at the surface.
The Nottingham effect is the heating or cooling of an emitting body due to the flow of electrons through its surface.
Each electron emitted from the conduction band of the metal surface is replaced by a Fermi level electron. 
The flow of electrons brings about a convective heat flux through the surface.
To see the direction of heat flux we look at Eq.~\eqref{QMcurrent}.
The integrand of \eqref{QMcurrent}, is used as a probability distribution function to calculate the average energy per electron leaving the surface.
Eq.~\eqref{Nottingham} describes the Nottingham flux $Q$, the heat flux escaping through the surface due to the emitted electrons.
If $\mu$ is the Fermi energy:
\begin{equation}
	Q(F,T)=\frac{1}{2\pi \hbar}\int_0^{\infty} (E-\mu) D(E)f(E)\ud E\quad[W/m^2].
	\label{Nottingham}
\end{equation}
The temperature at which the average energy of emitted electron matches the Fermi energy 
is called the inversion temperature, $T^*$.
The Nottingham effect causes heating of the surface for temperatures below the inversion temperature and cooling otherwise\cite{Charbonnier,Paulini}. 
We use the Nottingham heat flux as a boundary condition in our solution of the heat diffusion equation
to specify the gradient of temperature along the surface normal\cite{Kevin}.

\begin{figure}[ht]
	\includegraphics[width=0.4\textwidth]{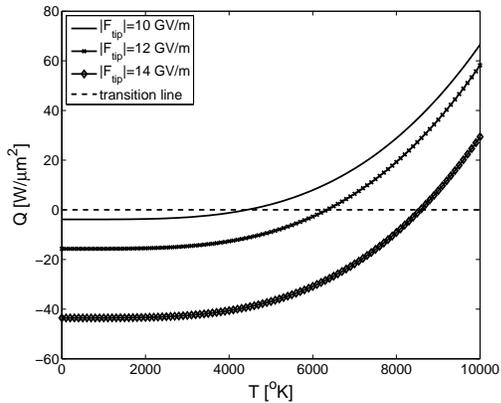}
	\caption{$Q$ the Nottingham heat flux leaving the system. 
Negative values indicate heat entering the system. $T^*$, the temperature at which heat flux is zero, 
is higher than melting temperature of copper for high electric fields.
\label{fig:not_transition}}
\end{figure}

Figure~\ref{fig:not_transition} shows the behavior of the Nottingham flux with respect to external electric field and temperature. 
Some immediate observations are the following.
i) The transition temperature is above the melting temperature, therefore the Nottingham effect is always a heating effect as far as the solid state is concerned. 
ii)The Nottingham flux is almost constant over the range of temperature below melting, which means heating is constant over time and no run away is possible due to the Nottingham effect. 
iii)The Nottingham flux falls sharply as the electric field decreases, which means, Nottingham effect contributes only at the tip. 

\section{Current Density in the Protrusion} \label{sec:u_solution}

Inside the protrusion we assume the current density and electric field are related by Ohm's Law $\mathbf{J}=\sigma \mathbf{E}$ where $\sigma$ is a spatially uniform electrical conductivity conductivity. 
The condition that charge density does not accumulate in time, $\nabla \cdot \mathbf{J}=-\partial \rho / \partial t = 0$ then yields a Laplace equation for the electrostatic potential.
Here we assume the dimensions of the protrusion are less than the skin depth.
We apply the following boundary conditions to the assumed azimuthally symmetric protrusion.
The radial current density at the axis is zero due to symmetry. 
The electric potential is assumed to be zero deep in the base supporting the protrusion.
Finally, the normal component of the electric field at the surface inside the protrusion is determined such that
the normal current density matches the emission current calculated in Eq.~\eqref{QMcurrent}.

\begin{figure}[ht]
		\includegraphics[width=0.4\textwidth]{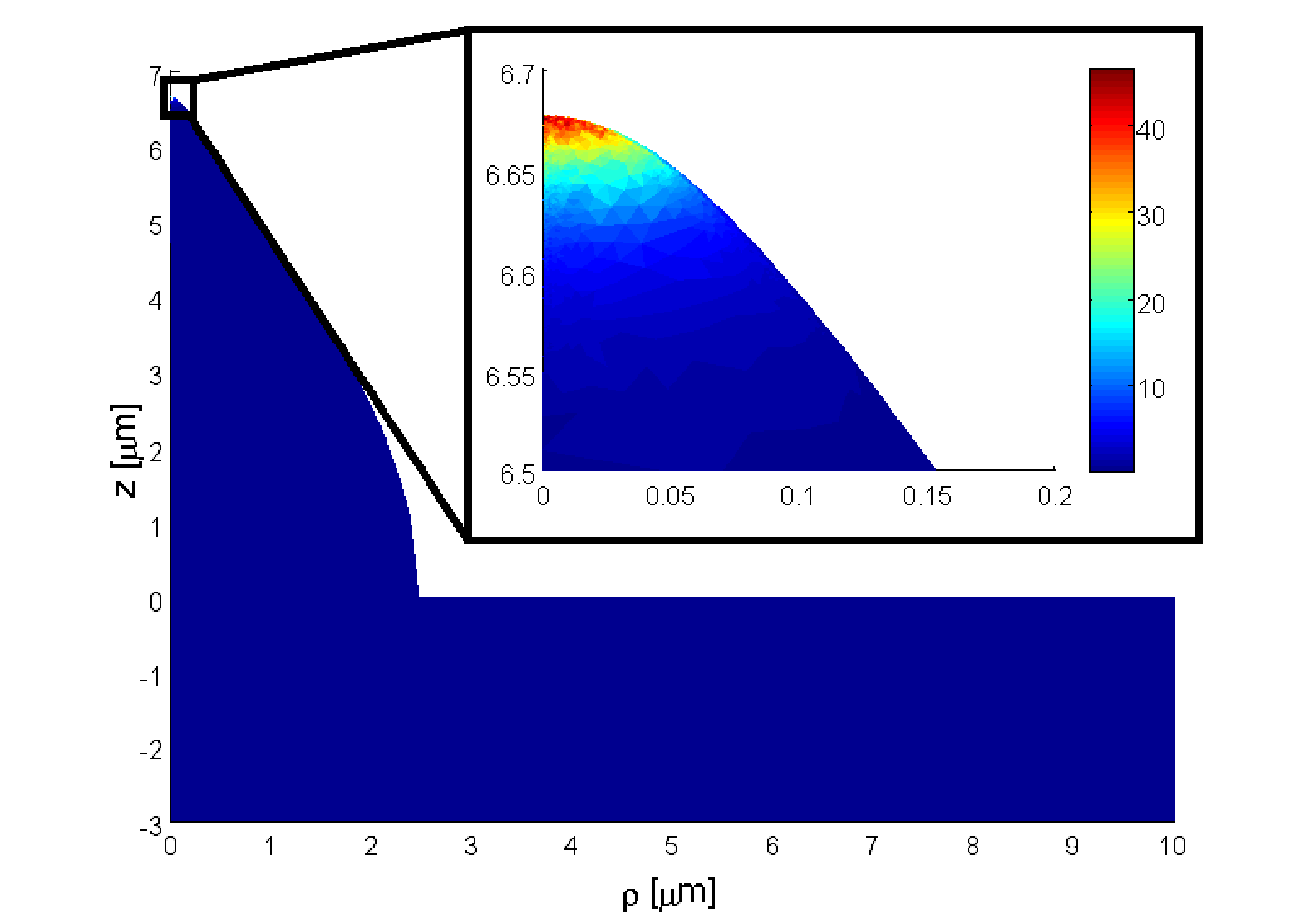}
		\caption{$|J|^2$ in $\lbrack A/{\mu m}^2 \rbrack^2$ for $a_o=1\:\mu m$, $ r=0.85$, $N=21$, $F_o=300\:MV/m$
	\label{fig:J_solution}}
\end{figure}

To solve the Laplace equation we use a finite element method (FEM) with triangular elements \cite{Sayas, Pepper}. 
The triangular grid is employed so that variations of the potential near the tip can be resolved accurately and efficiently.
The system is solved and the solution is plotted using MATLAB\textregistered\cite{matlab}.  
Fig.~\ref{fig:J_solution} shows the false color images of the square of the current density inside the metal.
The figure indicates that the Joule heating term is concentrated in the tip area. 
As an example, for a protrusion with $1\:\mu m$ base radius, half maximum of the heating term occurs at $12\:nm$ below the tip of the protrusion.

\section{Temperature rise in the protrusion} \label{sec:Temp_rise}

In the previous section, the current density inside the metal was obtained. 
Using this we calculate the volumetric Joule heating.
The temperature rise in the protrusion as a function of time, 
($\Delta T(r,z,t)$) is then obtained by solving the heat diffusion equation with the volumetric Joule heating term as a source and the following boundary conditions.
The radial temperature gradient on the axis is zero and the temperature rise is zero on the bottom of the base. 
The boundary condition on the protrusion surface is determined by the Nottingham heat flux defined in Eq.~\eqref{Nottingham}.
Specifically,  $\kappa \partial \Delta T / \partial x_n = Q$, where $\kappa$ is the thermal conductivity. 
The thermal conductivity, $kappa$, depends on the relaxation time of scattering thus the temperaure.
However, here, $\kappa$ is assumed to be temperature independent.
This assumption is validated both through simulations and another study by Jensen \emph{et. al.} \cite{JensenKappa}.
The space dependent part of the heat equation is calculated by using the finite element method. 
The resulting ordinary differential equations are solved in time using MATLAB\textregistered~ \textbf{ode15s}\cite{matlab}.

\begin{figure}[ht] 
	\includegraphics[width=0.4\textwidth]{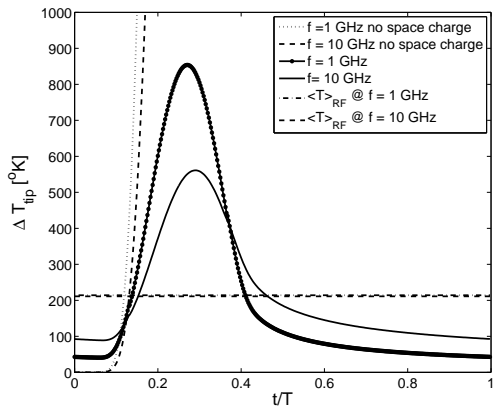}
	\caption{Comparison of $\Delta T$, change in tip temperature during the last RF cycle of a 100 ns pulse versus normalized time,
between 1 GHz and 10 GHz cases. $\beta \approx 47$ and $F_o = 300\:MV/m$. 
RF mean of the change in temperature is shown with the dashed line.
The dotted lines show the evolution of the tip temperature during the first cycle in the 1 GHz and 10 GHz cases, respectively, when space charge is neglected.
\label{fig:1cycle}}
\end{figure}

\begin{figure}[ht]
	\includegraphics[width=0.4\textwidth]{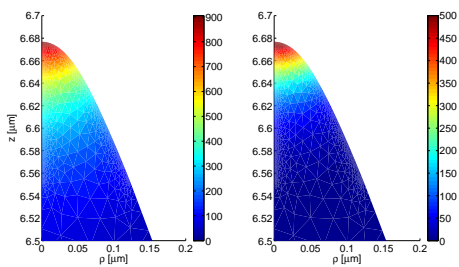}
	\caption{Spatial distribution of the temperature rise at the time of the peak excursion in the two cases, f=1 GHz on the left 
and f=10 GHz on the right.
\label{fig:Tempr}}
\end{figure}

The temporal variation of the RF electric field comes into play while solving these equations.
Specifically, both the Joule heating term and the Nottingham heat flux oscillate periodically in time with a period equal to that of the RF signal.
The result is that the temporal variation of the temperature rise, $\Delta T$, consists of a periodic variation superimposed on a slow temperature rise due to the average heating.
Figure~\ref{fig:1cycle} shows the temperature at the tip of a protrusion during the last cycle of 100 ns pulse for two different frequencies, $f=1\:GHz$ and $f = 10\:GHz$.
Other parameters are indicated in the caption, and time is normalized to the RF period.
The temporal pulse envelope of the RF in this case is a square wave.
As a consequence the temperature variation during cycle is nearly the same for all cycles in a pulse.
Thus, the maximum tip temperature is essentially determined during the course of a single cycle.
Also shown in Fig.~\ref{fig:1cycle} are the cycle averaged temperature rise, which is the same for both the 1 GHz and 10 GHz cases, 
and the temperature rise during the first cycle if space charge affects are not included in the model.
In this latter case the temperature quickly rises to the melting temperature $T_m=1357\:^oK$.
Thus space charge is an essential component of the model.
The maximum temperature excursion depends weakly on frequency, with the temperature rise in the 1 GHz case exceeding that of the 10 GHz case.
The trend with frequency is less dramatic than would be expected based on dimensional analysis of the one dimensional heat equation.
Specifically, if $T_{RF}$ is the period of the RF, one would expect the amount of heat deposited during a cycle, $U$ to scale as $U \sim T_{RF}$ for a given RF field amplitude.
In one dimension, the heat will diffuse a distance proportional to $\Delta \sim T_{RF}^{1/2}$.
Thus, the temperature rise, in one dimension, will scale as $U/\Delta \sim T_{RF}^{1/2}$.
The observed scaling with period is weaker. This is due to the pointed nature of the tip.
As the width of the protrusion increases with distance from the tip, the volume over which the heat is spread, $\Delta V$, increases with penetration depth.
We expect the temperature rise to scale as $\Delta T \sim U/\Delta V$, where for a conical tip $\Delta V \sim \Delta^3$.
Figure~\ref{fig:Tempr} shows the spatial distribution of the temperature rise at the time of the peak excursion in the two cases.
For these cases the ratio of temperature rise is $\Delta T_{1\:GHz} / \Delta T_{10\:GHz} = 1.70$.
Based on a conical tip we could expect  $\Delta T_{1\:GHz} / \Delta T_{10\:GHz} = 10/(1.7)^3 \approx 2.0$ 
which is consistent with the simulation.

To have a better understanding of the role of the Nottingham flux  we compare the volumetric equivalent of heating ($P_{N}$) caused by it, to the volumetric Joule heating $P_{J}=J^2/\sigma$.  
Assuming a semi-spherical tip of radius $r_T$, integrating the heat flux over the surface and dividing by the volume of the tip we find an equivalent volumetric heating rate $S_N=3 Q / r_T$ . 
As seen in Fig.~\ref{fig:not_transition}, $Q \approx 30\:W/{\mu m}^2$, for an enhancement factor of $\beta\approx44$ and background field of $F_o=300\:MV/m$.
Assuming a tip radius of $100\:nm$, we find $3 Q/r_T \approx 900 W/{\mu m}^3$.
We now take the ratio of the two volumetric heat sources: $P_{N}/P_{J}\sim 3\sigma Q/{r_T J^2}$. 
For the same value of $\beta$ background field and tip radius, we find $P_{J}=J^2/\sigma\approx13.5\:W/{\mu m}^3$ giving $P_{N}/P_{J}\approx66.6$.
This ratio decreases as temperature or electric field on the tip increases. 
When the range of our data is concerned, its minimum value is approximately 52. 
This implies that Nottingham heating dominates Joule heating at the tip. 
The dominance of Nottingham heating over the Joule heating was shown before also by Ancona in the treatment of thermal failure of field emitter tips \cite{ancona, Kevin}. 
The above order of magnitude calculation was validated by comparing the temperature at the tip with and without Joule heating; the difference is less than 0.5 \%.

\begin{figure}[ht]
	\includegraphics[width=0.4\textwidth]{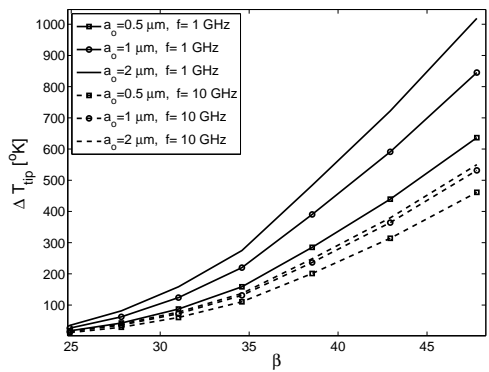}
	\caption{The maximum change in tip temperature versus $\beta$ for various values of scaling factor $a_o$.
Dashed lines describe the 10 GHz case.} 
\label{fig:T_compare}
\end{figure}

Figure~\ref{fig:T_compare} summarizes the dependence of the maximum change in tip temperature on various parameters, such as 
protrusion size, field enhancement factor and RF frequency.
Notice that, although the emission law is independent of RF frequency, high frequency causes less heat accumulation 
and hence lower swings of tip temperature.
Similarly, the quantity $a_o$ that determines the protrusion size as seen in eq.~\eqref{PCM_potential}, 
does not affect emission and the vacuum electric fields. 
However, it affects the tip temperature by modifying the space charge behavior.

\section{Conclusions}

\begin{figure}[ht]
	\includegraphics[width=0.4\textwidth]{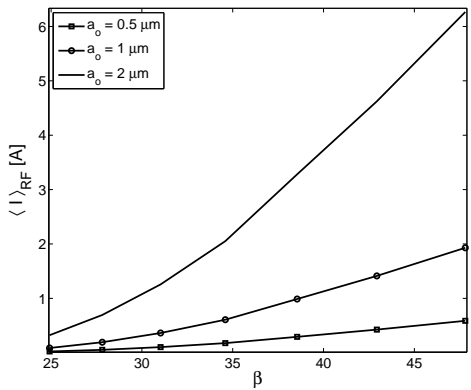}
	\caption{The RF averaged current versus $\beta$ for various values of scaling factor $a_o$.
\label{fig:current}}
\end{figure}

In this study we have performed extensive numerical simulations of electrical and thermal behavior of field emission on metal protrusions,
that are believed to be a source of dark current and a cause of breakdown in high-gradient accelerating structures. 
After unifying the thermal and field emission in the same numerical framework, we calculated the Nottingham and Joule heating terms and solved the heat equation to characterize the thermal evolution of emitters under RF electric field. 
Noteworthy conclusions can be listed as follows:

i)The contribution of Joule heating is negligible compared with the contribution of the Nottingham effect.
The Nottingham heating effect is approximately independent of temperature in the region of interest.
ii)The largest temperature rise occurs at the tip of the protrusion.
The temperature excursion is nearly periodic in times, with period given by the RF.
The peak temperature rise is several times the average rise, and the peak rise is relatively insensitive to RF period.
Due to both (i) and (ii), scenarios involving thermal run away are not plausible, 
unless some external phenomena that strongly couple heating to electric fields are present.
iii) Field emission is high enough to induce a space charge effect. 
The depression of electric field due to space charge causes a 5 to 7 fold reduction in emitted current.
This eliminates possibility of tip melting, including the case with the maximum emission allowed by the free electron model.
iv)The RF frequency effects the peak temperature. High frequency causes less temperature swing, therefore less peak temperature.
v)Both heating and emission are functions of field enhancement factor $\beta$. 
The geometric scaling factor $a_o$ slightly modifies heating and emission through the space charge mechanism.
Bigger protrusions emit more current in total. 
The RF averaged current scales with $a_o$ as seen in Fig.\ref{fig:current}.
vi)Time evolution of temperature and emission current is critically dependent on phenomena that
modifies the field enhancement factor. 
In this study, apart form the geometry, space charge played a decisive role. 
Gas discharge, plasma formation and ion bombardment mechanisms can be investigated by using our 
protrusion model, which is fairly simplified due to the afore mentioned considerations. 

\appendix
%    \numberwithin{equation}{section}
\section{The Space Charge Effect}

In this study we use a modified version of the Child Langmuir Law to account for electron space charge.
The Child Langmuir Law describes the steady current that flows from a planar cathode plate to a planar anode
a distance $x_D$ from the cathode at fixed potential $V_a$ when emission is space charge limited. 
The model includes the Poisson equation for the electrostatic potential $U(x)$.
\begin{equation}
	\frac{d^2}{dx^2} U(x)= \frac{|J|}{\epsilon_0 v(x)} = \alpha |J| U^{-1/2},\\
	\label{Child}
\end{equation}
along with boundary condition $U(x=0)=0$ and $U(x_D)=U_D$,
where, $J$ is the current density emitted from the surface of the cathode and $v(x)$  
is the velocity of electrons at point $x$. 
Energy conservation dictates:
$v(x)=(2 e U(x)/m)^{1/2}$ where $e$ is the fundamental charge and $m$ the electron mass.
Introducing the constant $\alpha = (m/(2 e))^{1/2}/\epsilon_o$ leads to the second equality in \eqref{Child}.
In the space charge limited case the value of current density is determined by the condition $dU/dx|_{x=0}$.
To incorporate emission one can use a relation of the form of \eqref{QMcurrent} where we set $J=J(F_S=dU/dx|_0,T)$.

Barbour \emph{et. al.} have done this by using the Fowler Nordheim emission and Child's equation together and provided a comparison with experiment\cite{Barbour}. 
For a historical review and a general mathematical formulation, the reader is encouraged to read the recent article by Forbes\cite{Forbes}.
%Jensen \emph{et. al.}  provides a quantum mechanical treatment and calculates transit times \cite{JensenSC}.
Numerical approaches are also employed to solve equilibrium equations for space charge limited flow for sharp edges and conical points of emitters\cite{antonsen}.

In our case, we have to address two issues to incorporate space charge in our model. 
First, we have a 3D emitter with nontrivial geometry, and second,
there is no well defined anode that fixes the domain and boundary condition as in \eqref{Child}.
We address these as follows:
Noting that the protrusion emits most of the current from the tip, we will ignore the emission from 
the rest of the protrusion. 
Moreover, for a highly curved tip, the electron current will spread radially as it flows away from the tip.
Consequently, the electron density will drop rapidly with distance along
the vertical axis of the protrusion away from the tip. 
Hence, the tip radius will serve as a scale 
with which we measure the region where the space charge is assumed to affect the emission. 
We will, thus, assume that one tip radius away from the tip,
the electric field reaches its vacuum value. 
This will serve as a boundary condition on the electric field, 
in the one dimensional equation.
The separation distance $x_D$, is now equal to the tip radius $r_T$.

In summary we solve Eq.~\eqref{Child}
with the boundary conditions
\begin{subequations}
	\begin{equation}
			U(0)=0,
	\end{equation}
	\begin{equation}
			\partial_x U|_{x_D}\approx F_V,\\
	\end{equation}
where $F_V$ is the vacuum field, and we define
	\begin{equation}
	 F_S=d U/d x|_0.
	\end{equation}
	\label{Keser}
\end{subequations}
The emission law is used to determine $|J|=J(F_S,T)$.
The solution of \eqref{Child} is standard. 
Writing $d(dU/dx)^2=2\alpha JU^{-1/2}dU$ and integrating, one finds:
$dU/dx=\sqrt{4\alpha JU^{1/2}+F_S^2}$. 
Integration from $x=0$ (where $U(0)=0$) to $x=r_T$ (where $d U/d x = F_V$)
yields
\begin{equation}
		12\alpha^2 J^2r_T = (F_V^3 - F_S^3) - 3 F_S^2(F_V-F_S).\label{solution}
\end{equation}
 which can be viewed as a transcendental equation for the field on the emitter ($F_S$) 
 once the emission law $J(F,T)$ is prescribed.
 
 The right hand side of \eqref{solution} is a monotonically decreasing function of $F_S$
and the left hand side is a monotonically increasing function of $F_S$.
 Thus we can expect a single solution satisfying $0<F_S<F_V$.
 Solution of the transcendental equation for three different tip radii $r_T$,
 and the generalized emission law Eq.~\eqref{QMcurrent} at room temperature is shown in Fig.~\ref{fig:SPalpha}.
 We note in the limit in which the emission current is large such that $F_S \ll F_V$,
 we recover a form of the Child-Langmuir  law, $J = F_V^{3/2}/\alpha(12 r_T)^{1/2}$.

\bibliography{Ref}

\end{document}